\documentclass[epj]{svjour}
\usepackage{graphicx}
\usepackage{color}
\usepackage{comment}
\usepackage[normalem]{ulem}
\newcommand{\lrangle}[1]{\langle{#1}\rangle}

\begin{document}
\title{Critical behavior of the contact process on small-world
networks}
\author{Ronan S. Ferreira \and Silvio C. Ferreira}
%
%
\institute{Departamento de F\'isica, Universidade Federal 
de Vi\c{c}osa, 36570-000, Vi\c{c}osa, Minas Gerais, Brazil}
\date{Received: date / Revised version: date}
%
\abstract{
We investigate the role of clustering on the critical behavior of 
the contact process (CP) on small-world networks using the Watts-Strogatz 
(WS) network model with an edge rewiring probability $p$. The critical 
point is well {predicted} by a homogeneous cluster-approximation for 
the limit of vanishing clustering ($p\rightarrow 1$). The critical 
exponents and dimensionless moment ratios of the CP are
in agreement with those predicted by the mean-field theory for any $p>0$.
This independence on the network clustering shows that the small-world
property is a sufficient condition for the mean-field theory to correctly
predict the universality of the model. Moreover, we {compare}
the CP dynamics on WS networks with rewiring probability $p=1$ and random
regular networks and {show} that the weak heterogeneity of the WS
network slightly changes the critical point but does not alter other
critical quantities of the model. 
} 
\maketitle

\section{Introduction}
\label{sec:intro} The Watts-Strogatz (WS) model is recognized as a typical
example of a small-world network~\cite{watts1998collective}. The small-world
property in a network constituted by vertices connected by edges means that the
average shortest path between vertices, $\lrangle{l}$, increases
logarithmically or slower with the number of vertices $N$. This property is a
central feature shared by many complex networks~\cite{newman2010networks,Reka}.
Differently from scale-free (SF) networks, which do not present a characteristic
scale for the fluctuations around the average number of connections (the vertex
degree) $\lrangle{k}$, the WS networks have a {narrow} degree distribution
$P(k)$, defined as the probability that a randomly chosen vertex has degree $k$,
which {decays} exponentially fast for large $k$. These networks are commonly
treated as homogeneous~\cite{Vespignani63,pastor2001epidemic} even being
heterogeneous in a strict sense.

In addition to the small-world property, a high average clustering 
coefficient $\lrangle{c}$ is another important topological property 
exhibited by several complex systems~\cite{newman2010networks}. This 
quantity is defined by~\cite{Barrat}
\begin{equation}
\lrangle{c}=\frac{1}{N}\sum_i\frac{e_i}{k_i(k_i-1)/2},
\end{equation}
where $e_i$ is the number of edges between the neighbors of a vertex $i$
and the denominator contains the maximum number of these edges in a 
vertex with $k_i$ connections.

In the WS model, the clustering is controlled by the rewiring probability
$p$ such that {this} model interpolates from a {regular lattice} for 
$p=0$ to an almost random network for $p=1$. Despite of the high 
clustering coefficient observed in many real-world networks presenting a 
long tailed degree distribution $P(k)\sim k^{-\gamma}$ ($2<\gamma<3$), 
random SF networks have a vanishing clustering in the {thermodynamic}
limit~\cite{catanzaro2005generation,colomer2012clustering}. Contrastingly,
both small-world and high clustering properties can coexist {in} WS
networks, providing an interesting substrate to study dynamical
processes~\cite{Vespignani63,barrat00,PhysRevE.61.5678,zanette01}.

Simple dynamical processes with absorbing states exhibit rich features
when running on the top of SF
networks~\cite{castellano06,Hong98,cp_quenched,sisFerreira}. The simplest
example is the contact process (CP)~\cite{harris74}, a paradigmatic      
interacting particle system involving spontaneous annihilation of
particles at {unitary} rate and catalytic creation in a pair of 
occupied/empty nearest neighbors at a rate $\lambda$. The configuration 
devoid from {particles} is called absorbing: once the system  
visited an absorbing configuration the dynamics remains permanently 
trapped within this state. The model undergoes an absorbing state phase 
transition for a critical value $\lambda_c$ of the control 
parameter~\cite{marro99}.

The CP dynamics has been investigated in networks using the heterogeneous
mean-field (HMF)
theory~\cite{castellano08,boguna2009langevin,cp_annealed}, in which all
dynamical correlations are disregarded and quantities of interest depend
only on the vertex degree~\cite{dorogov}. {According to} the HMF 
theory for the CP, the critical density of particles $\rho$ and 
characteristic time $\tau$ obey the finite-size scaling (FSS)
relations~\cite{boguna2009langevin}
\[\rho\sim(gN)^{-1/2} \mbox{~~and~~} \tau\sim (N/g)^{1/2},\] where the
factor $g=\lrangle{k^2}/\lrangle{k}^2$ depends on the form of $P(k)$ and
introduces an anomalous dependence on the cutoff $k_c$ of the distribution
$P(k)$ for {the case of} SF networks~\cite{boguna2009langevin}. 
Even for quenched SF networks having an asymptotically null clustering 
coefficient, the HMF theory is able to predict the correct FSS exponents 
of the CP dynamics independently of the presence of dynamical
correlations~\cite{cp_quenched}. However, dynamical correlations lead to
wrong predictions of the critical points.

While the CP dynamics on nonclustered SF networks has been subject of
several recent
studies~\cite{castellano06,Hong98,cp_quenched,castellano08}, the
role of the clustering has not been carefully investigated up to this
moment. In this work, we address this issue by investigating the
criticality of the CP model on the top of WS networks with different rewiring
probabilities. Our results show that the critical exponents agree with
those of the HMF theory even for highly clustered networks with
$p>0$. On the other hand, {the position of the critical point} 
strongly depends on the parameter $p$. In addition, we compare the WS 
results with those for nonclustered homogeneous networks obtained with 
the random regular (RR) networks~\cite{sisFerreira}.

We have organized this paper as follows. The model is presented in
section~\ref{sec:models}. Section~\ref{sec:mf} is devoted to a brief
review of the critical properties of the CP in a mean-field level.
Simulations are presented and discussed in section~\ref{sec:results}. Our
conclusions and remarks are drawn in section~\ref{sec:conclusions}. Two
appendixes with pair and three-vertex approximations for the CP 
complement the
paper. These approximations, particularly that for three-vertices, 
can be useful in the investigation of other similar problems.

\section{The model}
\label{sec:models} 

The CP is a reaction-diffusion system including self-anni\-hilation and
catalytic creation of particles in pairs of occupied and empty
vertices~\cite{harris74}. The CP rules in arbitrary graphs are defined as
follows~\cite{marro99}: A vertex $i$ can be occupied ($\sigma_i=1$) or empty
($\sigma_i=0$). At a rate $\lambda$, occupied vertices try to create an
offspring in one of their nearest neighbors selected at random. Creation events
successes only on empty vertices. An occupied vertex becomes spontaneously empty
at unitary rate while an empty vertex $i$ is occupied at a rate $\lambda \sum_j
\sigma_j/k_j$ where the sum runs over all neighbors of $i$ and  $k_j$ 
is the respective vertex degree. This dynamics can drive the system to a 
frozen phase devoid from particles, the absorbing state. However, if the 
control parameter
$\lambda$ is sufficiently large the active phase becomes stable and a finite
fraction of the network is occupied for $t\rightarrow\infty$. The stationary
density of occupied vertices $\rho$ is the order parameter. The critical value
$\lambda_c$ separates the active ($\rho>0$ for $\lambda>\lambda_c$) and
absorbing ($\rho = 0$ for $\lambda\le\lambda_c$) phases.

We used the original WS model~\cite{watts1998collective} as the underlying
substrate for the CP dynamics. This model allows to interpolate from a {regular
lattice} to a random network combining two important topological features: a
high clustering and a low average shortest path between vertices. According to
the WS model~\cite{watts1998collective}, a small-world network can be built from
$N$ vertices initially ordered in a {one-dimensional lattice} with periodic
boundary conditions, in which each vertex has $K$ connections with the nearest
neighbors. Vertices are clockwise visited and for each vertex the
clockwise edges  are rewired with probability $p$, also in the clockwise sense.
The rewiring rules generate connected networks and conserve the
number of edges implying that $\lrangle{k}=K$. We used $K=6$ in our
simulations.

The rewiring of edges introduces $pNK/2$ long-range connections, reducing the
average shortest path $\lrangle{l}$ between vertices. For $p=0$, the network is
a {regular lattice} with $\lrangle{l}\approx N/2K$ and a high average
clustering coefficient $\lrangle{c}\approx3/5$~\cite{Reka}. On the other hand,
for $p\rightarrow1$, the network converges to both vanishing clustering
$\lrangle{c}\sim K/N$ and small average shortest path
$\lrangle{l}\sim\ln(N)/\ln(K)$, typical of random networks~\cite{Reka}.
Interestingly, there is a broad range of $p$ values where a large $\lrangle{c}$
and a short $\lrangle{l}$ are concomitant. However, for any finite $p$ and
sufficiently large sizes, the presence of shortcuts in the networks renders the
small-world property~\cite{barrat00,barthe99}. 

\section{Mean field approximation for the CP on networks}
\label{sec:mf}
Analytical insights for the critical behavior of the CP on random
networks can be obtained from the HMF
theory~\cite{castellano06,castellano08,boguna2009langevin,cp_annealed}.
Assuming networks with degree distribution $P(k)$ and the absence of
dynamical and degree correlations, the HMF equation for $\rho_k$
is~\cite{castellano06}:
\begin{equation}
 \frac{d\rho_{k}}{dt}=-\rho_{k}+\frac{\lambda
k}{\lrangle{k}}(1-\rho_{k})\rho,
\label{eq:HMF} 
\end{equation}
where 
\[
\rho=\sum_kP(k)\rho_k 
\]
is the overall density of particles in the networks. The WS network is
asymptotically free from degree correlations due to the fast decay of the
$P(k)$ for large $k$. It was shown that the critical point for
Eq.~(\ref{eq:HMF}) is $\lambda_c=1$ independently of the degree
distribution~\cite{boguna2009langevin}. 

In the stationary state, $d\rho(t)/dt=0$, we have
\begin{equation}
\rho_{k}=\frac{\lambda k\rho/\lrangle{k}}{1+\lambda k\rho/\lrangle{k}}.
\label{eq:rhok}
\end{equation}
Near the critical point, such that $\rho k \ll 1$, Eq.~(\ref{eq:rhok})
yields $\rho_k\simeq \lambda k\rho/\lrangle{k}$. Plugging this result in
Eq.~(\ref{eq:HMF}) we obtain
\begin{equation}
 \frac{d\rho}{dt}=- \rho +\lambda(1-g\rho)\rho,
\label{eq:rhog} 
\end{equation}
where $g=\lrangle{k^2}/\lrangle{k}^2$ carries the dependence on the degree
distribution $P(k)$. For the WS networks the factor $g$ is practically
independent of $N$. Therefore, we found the usual mean-field result for CP
where a continuous phase transition $\rho \sim (\lambda-\lambda_c)^\beta$
with $\lambda_c=1$ and $\beta=1$ are found.

In a mean-field level, equation~(\ref{eq:rhog}) is the macroscopic
equation of the one-step process~\cite{castellano08}:
\begin{equation}
\begin{array}{lll}
w_{n-1,n} & = & n\\
w_{n+1,n} & = & n(1-gn/N)
\end{array},
\label{eq:w}
\end{equation}
where $w_{n,m}$ is the transition rate from the state $m$ to the state 
$n$.
For finite size systems, one must consider the quasistationary (QS)
analysis where only surviving samples are considered in the
averages~\cite{marro99}. The QS analysis of this one-step processes was
performed in Ref.~\cite{cp_annealed}. The central result is the
probability $\bar{P}_n$ that $n$ vertices are occupied in the QS regime is
given by a scaling form 
\begin{equation}
 \bar{P}_n=\frac{1}{\sqrt{\Omega}}f\left(\frac{n}{\sqrt{\Omega}}\right),
\label{eq:pn}
\end{equation}
where $\Omega=N/g$ and $f(x)$ is a scaling function. One can directly
derive the scaling of the basic critical properties of the system
as~\cite{cp_annealed}
\begin{equation}
 \bar{\rho}=\frac{1}{N}\sum_n n\bar{P}_n\sim (gN)^{-1/2}\sim N^{-1/2}
\label{eq.6}
\end{equation}
and 
\begin{equation}
 \tau=\frac{1}{\bar{P}_1}\sim (N/g)^{1/2}\sim N^{1/2}.
\label{eq.7}
\end{equation}

The network heterogeneity still plays a role in the supercritical 
phase. The density of particles presents a dependence on the degree
distribution given by~\cite{boguna2009langevin}
\begin{equation}
\bar{\rho}\sim \frac{\bar{\rho}_{hom}}{g},
 \label{eq:rhosup}
\end{equation}
where $\bar{\rho}_{hom}$ is the solution for the strictly homogeneous 
networks with $g\equiv 1$. {Equation~(\ref{eq:rhosup}) can be
directly obtained either from Eq.~(\ref{eq:rhog}) of this paper or Eq. 
~(57) of Ref.~\cite{boguna2009langevin}}. 

\section{Results}
\label{sec:results}
The CP simulations were performed with the usual scheme \cite{marro99}: An
occupied vertex $i$ is chosen at random and the time incremented by
$1/[(1+\lambda)n(t)]$, where $n(t)$ is the number of particles at time $t$. With
probability $p=1/(1+\lambda)$, the particle $i$ is eliminated. With the
complementary probability $1-p$, one of the $k_i$ nearest neighbors of $i$
is randomly chosen and, if empty, occupied. The finite size inherent to
simulations will force the system to visit the absorbing configuration
even for $\lambda>\lambda_c$ at some sufficiently long time due to the
stochastic fluctuations~\cite{marro99}. Therefore, a suitable simulation
method is necessary to circumvent this problem.

We adopted a QS simulation method~\cite{de2005simulate} (see
Refs.~\cite{cp_quenched,sisFerreira,cp_annealed,sander2013} for applications of
the QS method in dynamical process on networks), in which the system history is
used to replace the absorbing state. The method is implemented by keeping and
constantly updating a list of $M=400$ active configurations taken from the
evolution of the system. Every time the system visits an absorbing state, this state
is replaced by a configuration picked up at random from the list. An update of
the list consists in replacing a stored configuration, chosen at random, by the
current one with probability $p_{rep}$. In our simulations we used
$p_{rep}=10^{-2}$ to $10^{-4}$ (The large the network size the smaller the value
of $p_{rep}$). After a relaxation time $t_r=10^6$, the probability $\bar{P}_n$
is determined during an averaging time $t_a=10^7$. All relevant quantities
derive from the QS distribution $\bar{P}_n$. The QS density of particles 
given by Eq.~(\ref{eq.6}) and the characteristic time given by 
Eq.~(\ref{eq.7}), are the basic quantities that we use to compute 
critical 
exponents. 

Since HMF theory disregards dynamical correlations between nearest
neighbors, one does not expect that the mean-field critical point
$\lambda_c=1$ is the correct value for quenched
networks~\cite{cp_quenched,munoz2010griffiths}. The critical point of the
CP on networks was determined using moment ratios of the order
parameter~\cite{cp_quenched,Dickman_moments} that are independent of
the network size when $\bar{P}_n$ has a scaling form as in
Eq.~(\ref{eq:pn}). This method was successfully applied to find the
thresholds of the dynamical processes taking place on SF
networks~\cite{cp_quenched,sander2013}. The moment ratios in the form
\begin{equation}
M^q_{rs}=\frac{\lrangle{\rho^q}}{\lrangle{\rho^r}\lrangle{\rho^s}}
\textrm{ with }q=r+s,
\label{eq:moms}
\end{equation}
intersect at $\lambda_c$ for different network sizes~\cite{cp_quenched}.

\begin{figure}[ht]
\centering
\includegraphics[width=8.5cm]{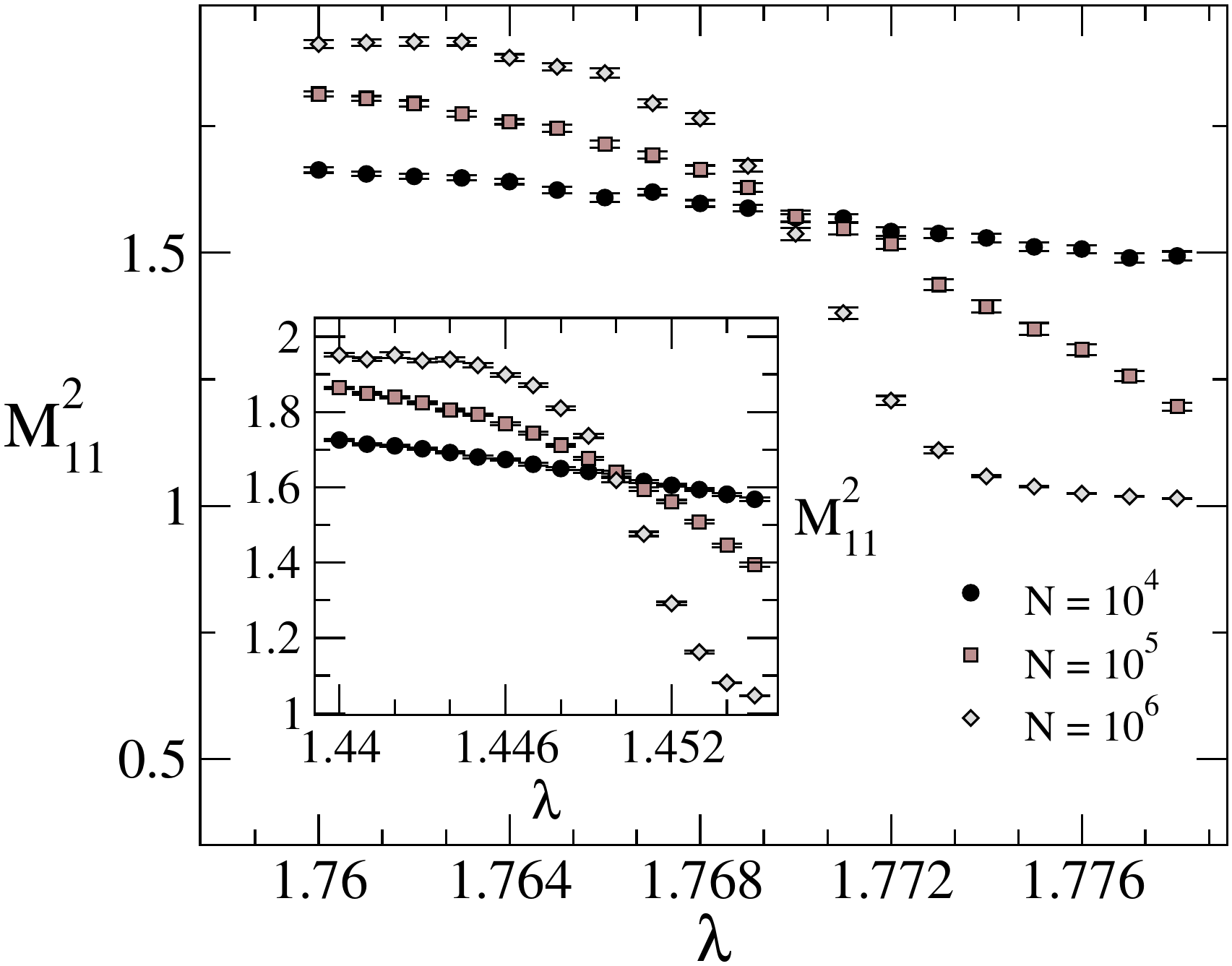}
\caption{Second order moment ratio for the CP dynamics on WS networks
with $\lrangle{k}=6$. The main plot shows the curves for the rewiring
probability $p=0.01$ and the inset for $p=0.10$.}
\label{fig:moms001}
\end{figure}

\begin{figure}[ht]
\centering
\includegraphics[width=8.5cm]{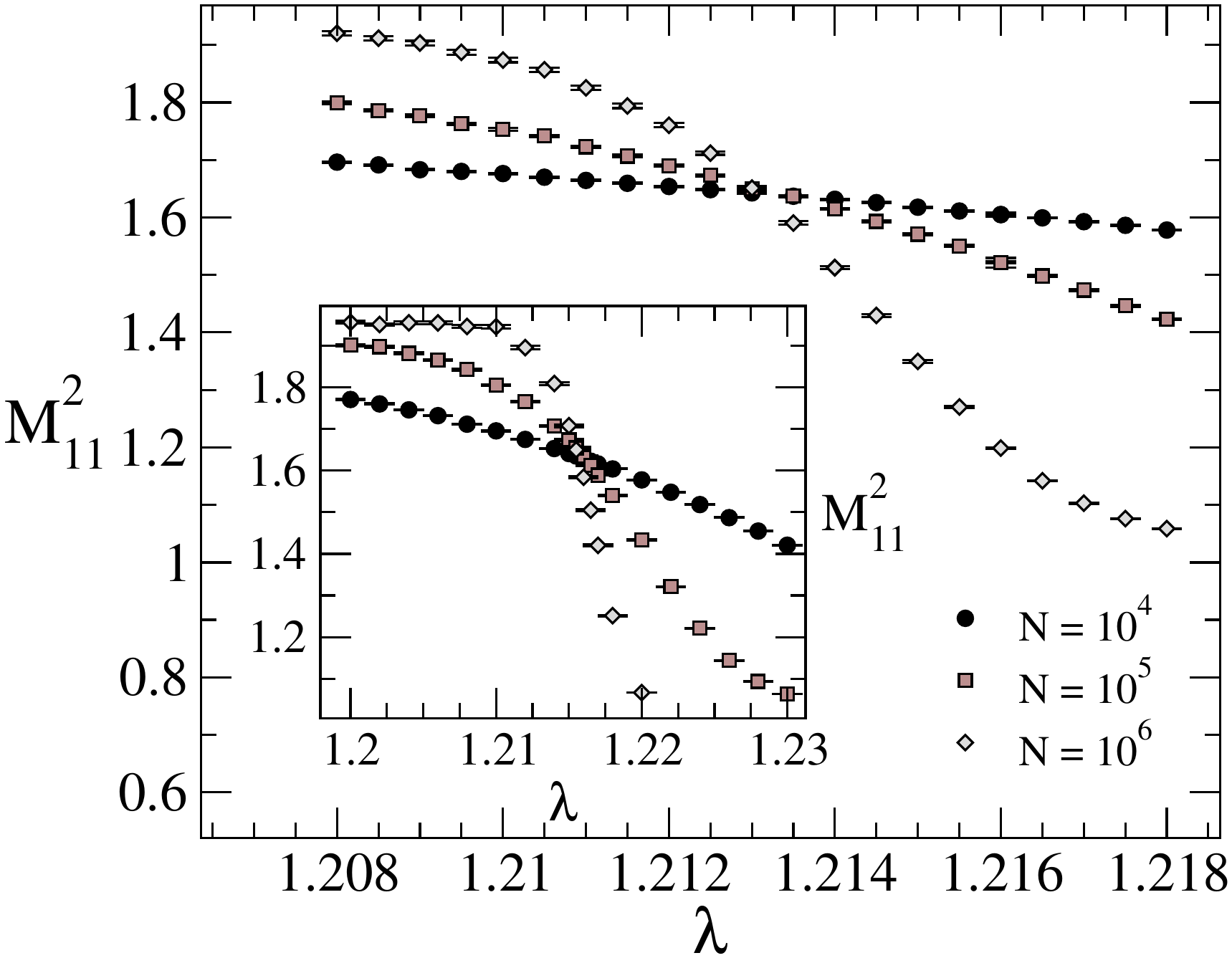}
\caption{Main plot: second order moment ratio for the CP dynamics on WS
networks with $p=1.00$ and $\lrangle{k}=6$. Inset: the same analysis for
random regular networks with fixed degree $k=6$.}
\label{fig:momWSRRN}
\end{figure}

We investigate the CP dynamics on networks with different clustering
coefficients built with rewiring probabilities $p=0.01$, $p=0.10$, and
$p=1.00$. The determination of the critical points for $p=0.01$ and
$p=0.1$ are exemplified in Fig.~\ref{fig:moms001}, in which we show the
moment ratio $M^2_{11}=\lrangle{\rho^2}/\lrangle{\rho}^2$ against
creation rate. The corresponding curves for $p=1$ are shown in
Fig.~\ref{fig:momWSRRN}. The same crossing points were obtained for higher
order moment ratios. The critical points and critical moment ratios up to
fourth order are reported in Table~\ref{tab:mom}.

\begin{table}[ht]
\centering
 \begin{tabular}{ccccc}\hline\hline
 Network&$\lambda_c$&$M_{11}^2$&$M_{21}^3$&$M_{22}^4$\\\hline
  $p=$0.01&1.7691(1)&1.64(4)&2.1(1)&3.2(2)\\
  $p=$0.10&1.4498(1)&1.65(3)&2.1(1)&3.4(1)\\
  $p=$1.00&1.2130(1)&1.65(3)&2.1(1)&3.4(1)\\
  RR&1.2155(1) &1.664(6)&2.18(1)&3.41(3)\\
  ANN&1&1.667(3)&2.190(4)&3.452(3)\\\hline\hline
 \end{tabular}
\caption{Critical points $\lambda_c$ and critical moment ratios for CP on 
WS
networks with different rewiring probabilities and random regular (RR) 
networks
with $\lrangle{k}=6$. The values corresponding to annealed networks (ANN) 
taken
from Ref.~\cite{cp_quenched} are also included for
comparison.}
\label{tab:mom}
\end{table}

Despite of the slower convergence observed for small values of the
rewiring probability $p$, the moment ratios are not affected by dynamical
correlations between vertices and agree, within the numerical
uncertainties, with the values found for annealed networks, which are
independent of the degree distribution $P(k)$~\cite{cp_quenched}. 
However, these correlations displace the position of the critical point
in relation to the HMF prediction $\lambda_c=1$. Notice that the
displacement increases for highly clustered networks. Moreover, the
threshold $\lambda_c^{PA}=1.20$ obtained in a standard pair-approximation
for nonclustered networks~(see appendix~\ref{app:pair}), which fits
remarkably well critical points of CP and related models in random SF
networks~\cite{cp_annealed,sander2013,munoz2010griffiths}, is a good
approximation only in the limit of $p\rightarrow 1$ when the network
exhibits an asymptotically vanishing clustering
coefficient~\cite{barrat00}. So, a necessary condition for the accuracy of
the thresholds in a pair-approximation is a low clustering.
The accuracy is substantially enhanced when compared with a three-vertices
approximation $\lambda_c^{TA}=1.21103$ (see Appendix~\ref{app:three}) 
for
which the difference is of only 0.2\%. A simple modification of the
pair-approximation including the possibility of loops, yields
$\lambda_c=k/(k-1-c/2)$ where $c$ is the clustering coefficient. Details
of the theory are given in appendix~\ref{app:pair}. This result
qualitatively explains the decay of $\lambda_c$ for increasing 
$p$. However, this approach is still quantitatively 
inaccurate for small values of $p$, as shown in appendix~\ref{app:pair}.

\begin{figure}[ht]
\centering
\includegraphics[width=8.5cm]{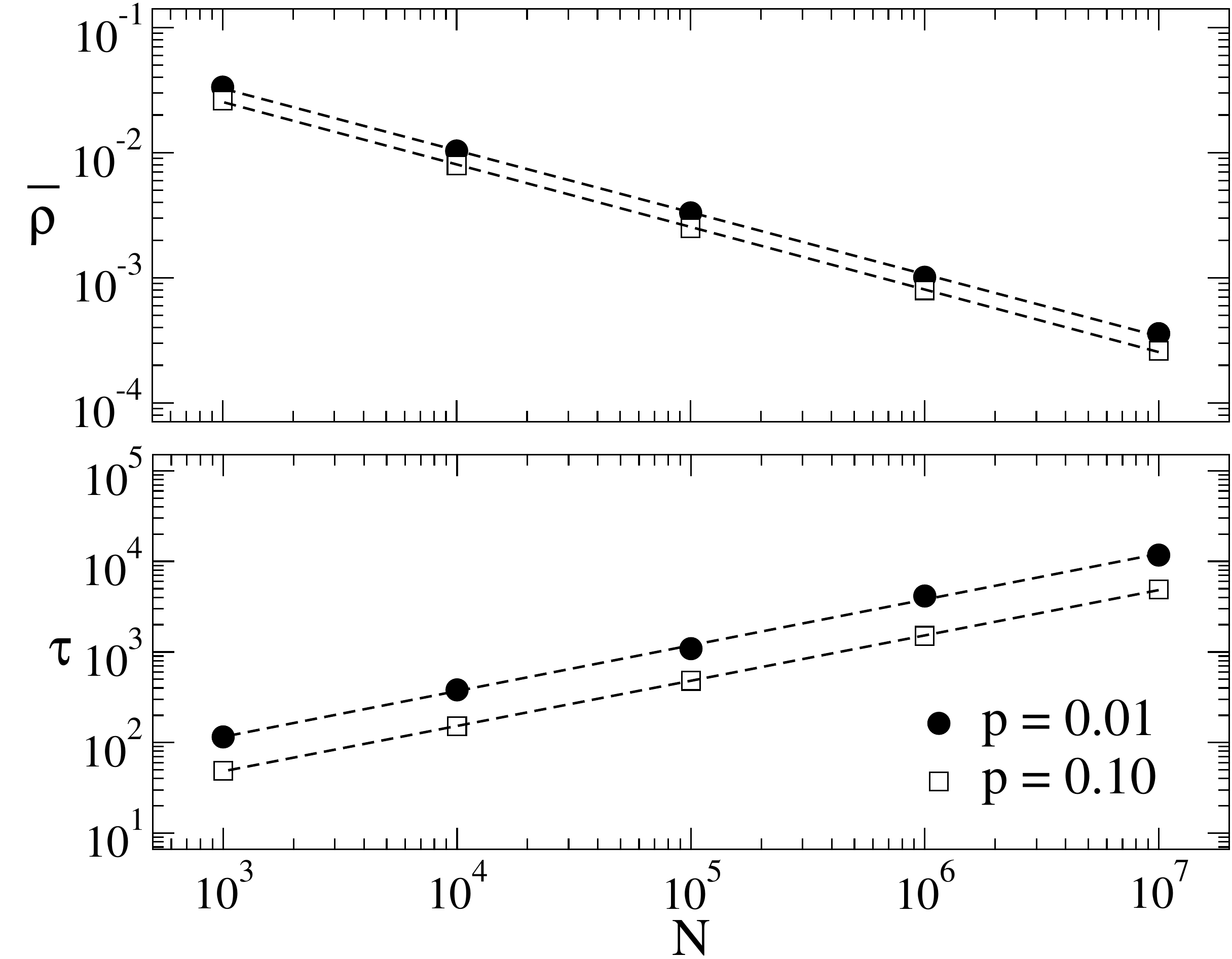}
\caption{FSS of relevant quantities for the CP on highly clustered WS
networks with $\lrangle{k}=6$ at the critical point.
The QS density of particles and characteristic time against network size
are shown in top and bottom panels, respectively. Lines are power law
regressions used to determine $\hat{\nu}$ and $\hat{\alpha}$.}
\label{fig:rhotau001}
\end{figure}

FSS analyses were performed for networks sizes ranging from $10^3$ 
to $10^7$. Figure~\ref{fig:rhotau001} shows the scaling of $\bar{\rho}$ 
and $\tau$ as functions of $N$ at the critical point. The results for
highly clustered structures ($p=0.01$ and $p=0.10$) show that both
quantities obey the scaling exponents of HMF theory (see
Table~\ref{tab:expo}). An interesting remark is that both FSS exponents as
well as moment ratios of CP on WS networks agree with the corresponding
values on annealed network. For highly heterogeneous SF networks, HMF
theory yields exponents in good agreement with simulations but fails in
predicting moment ratios dependent of $P(k)$~\cite{cp_quenched}. Finally,
corrections to the scaling, which play a central role for critical CP in
highly heterogeneous
substrates~\cite{cp_quenched,boguna2009langevin,cp_annealed}, are absent
in WS networks.
\begin{table}[ht]
\centering
 \begin{tabular}{cccc}\hline\hline
   Network~&$\hat{\nu}$&$\hat{\alpha}$&MF\\\hline
  $p=$0.01 ~&0.50(1) ~&0.49(2) ~&1/2\\
  $p=$0.10 ~&0.50(1) ~&0.49(1) ~&1/2\\
  $p=$1.00 ~&0.50(1) ~&0.50(2) ~&1/2\\
  RR ~&0.50(1)  ~&0.50(2) ~&1/2\\\hline\hline
 \end{tabular}
\caption{FSS exponents $\bar{\rho}\sim N^{-\hat{\nu}}$ and $\tau\sim 
N^{\hat{\alpha}}$ for the CP on WS and random regular (RR) networks with 
$\lrangle{k}=6$. The mean-field result is also shown.}
\label{tab:expo}
\end{table}

\begin{figure}[ht]
\centering
\includegraphics[width=8.5cm]{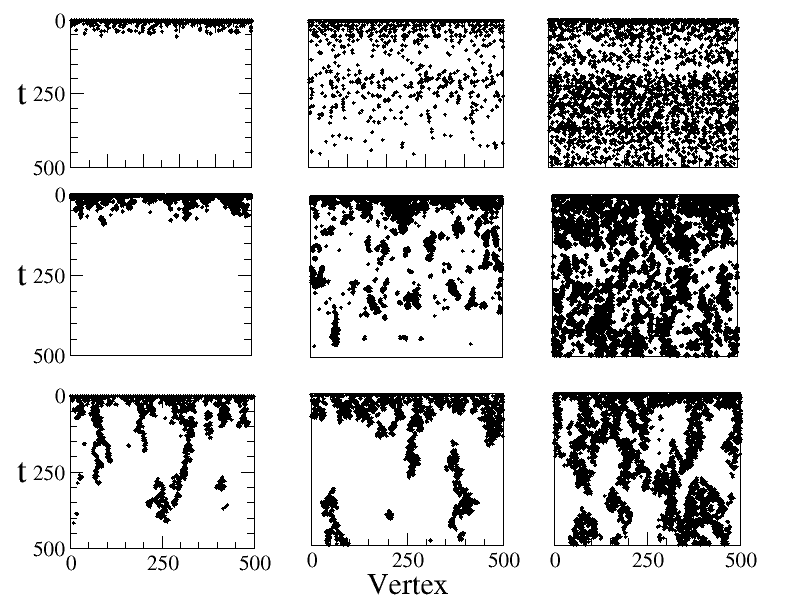}
\caption{Space-time patterns generated by the CP dynamics on WS networks
starting with all vertices occupied. Bottom panels correspond to a highly
clustered network with rewiring probability $p=0.01$, middle to $p=0.1$
while the top considers a nonclustered network with $p=1.0$. Subcritical
(left), critical (center), and supercritical (right) configurations are
shown.}
\label{fig:evolpcws}
\end{figure}

The role of the clustering in the evolution of the CP on WS networks is
qualitatively illustrated in Fig.~\ref{fig:evolpcws}, in which we show
space-time patterns consisting of configurations taken between time 
intervals
$\Delta t=1$, sequentially ordered in time. For highly clustered networks
($p=0.01$), one can neatly see localized activity lasting for a finite 
time. The
few shortcuts are responsible by the emergence of isolated patches of 
activity
that are forbidden in {regular lattices}~\cite{henkel2009}. For $p=0.1$, 
the
clustered patterns are highly reduced while the low clustered network 
$p=1.0$
does not exhibit space-time structures due to high number of shortcuts.

\begin{figure}[ht]
\centering
\includegraphics[width=8.5cm]{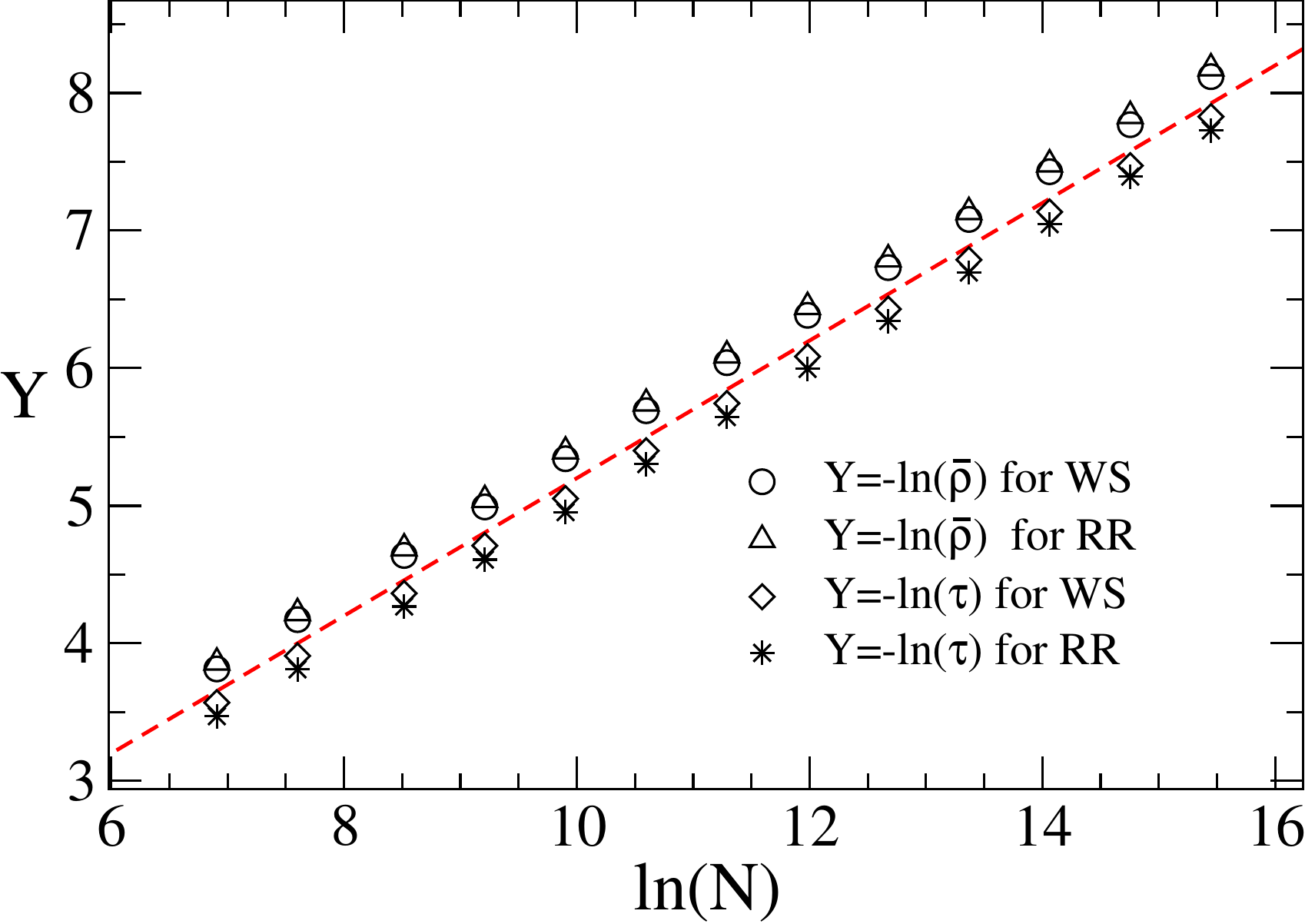}
\caption{Double logarithmic plots of the QS quantities of the critical CP
on
WS and RR networks with $\lrangle{k}=6$. Dashed line has a slope
1/2.}
\label{fig:rhotauwsrrn}
\end{figure}

The weak heterogeneity of WS network does not affect the universality of
neither critical moment ratios nor exponents, which agree with the
HMF theory. However, WS network does not have a
strictly homogeneous degree distribution. In order to investigated the
role of this weak heterogeneity, we compare the CP dynamics on WS
networks with $p=1$ (the most randomized case) with that on RR
networks. In the latter, all vertices have exactly the
same degree but connections are done at random excluding multiple and
self-connections~\cite{sisFerreira}. The critical points are slightly
different as one can see in Table~\ref{tab:mom}. Curiously, the critical
point for the strictly homogeneous RR networks is further from mean-field
critical point ($\lambda_c=1.21103$ in a three-site approximation) than
WS networks with $p=1$. We expect that higher order mean-field theories
must converge to the critical point observed for the homogeneous RR
networks since heterogeneity is expected to play some role in dynamical
correlations and consequently in the critical point.
Figure~\ref{fig:rhotauwsrrn} shows the FSS of the critical quantities
$\bar{\rho}$ and $\tau$. As expected, the critical exponents and moment
ratios for CP on RR networks agree very well with the mean-field theory as
shown in tables~\ref{tab:mom} and \ref{tab:expo}.

\begin{figure}[ht]
\centering
\includegraphics[width=8.5cm]{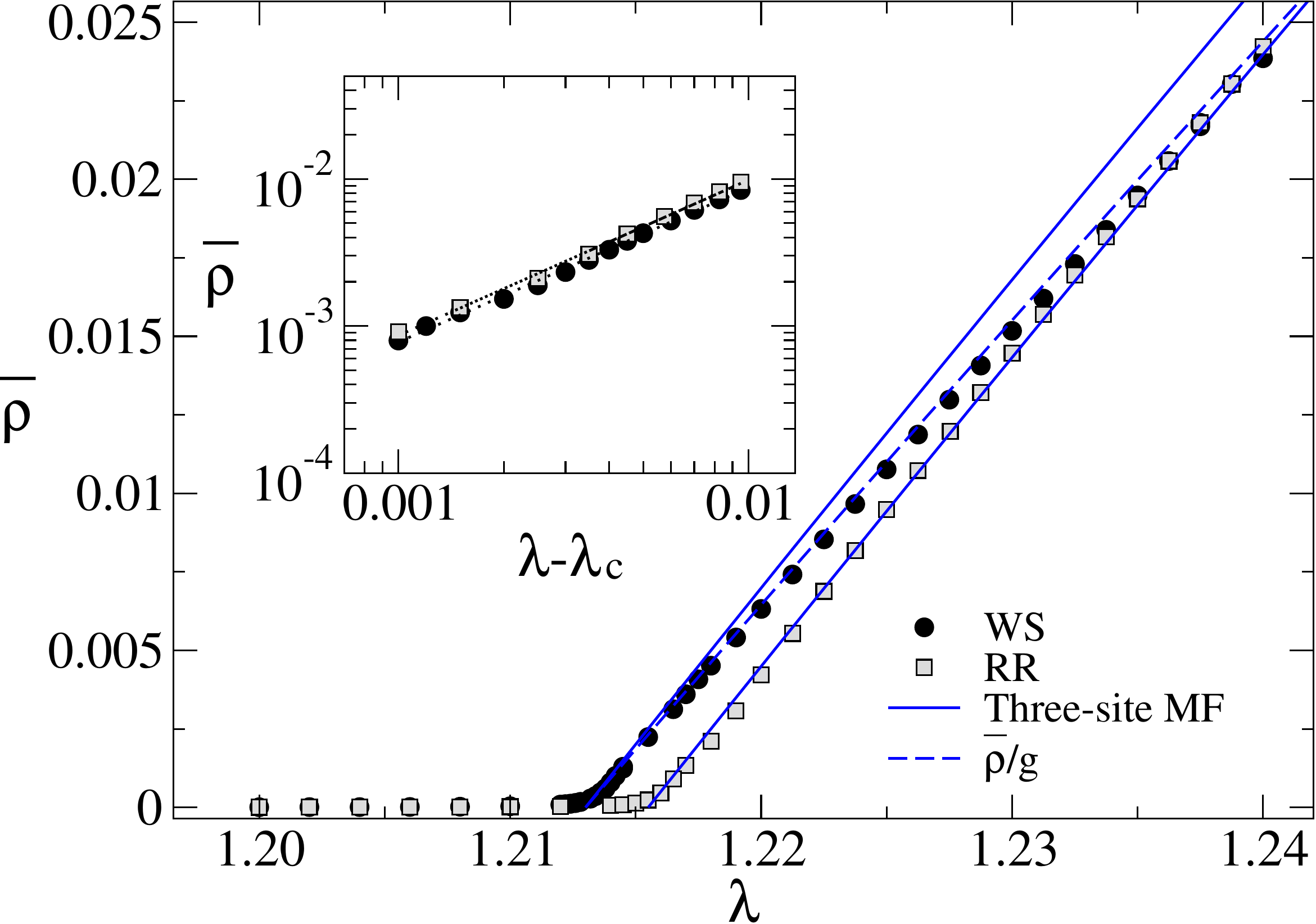}
\caption{QS density against the creation rate for CP dynamics on WS
networks with $p=1$ and RR networks, both with $N=10^7$ vertices. Solid
lines represent the mean-field solution (See appendix~\ref{app:three})
shifted to vanish at the actual critical points given in
Table~\ref{tab:mom}. Dashed line shows the solution predicted by
heterogeneous mean-field theory, Eq.~(\ref{eq:rhosup}). Inset shows the
supercritical density against the distance to critical point.}
\label{fig:rholamblambc}
\end{figure}

Further comparisons between CP dynamics on WS and RR networks are given in
Fig.~\ref{fig:rholamblambc}, where QS density is shown as a function of
the creation rate. The critical exponent $\beta=1.05(4)$ agrees with the
mean-field exponent $\beta=1$. The density predicted by the mean-field
theory, Eqs.~(\ref{eq:rhopair}) or (\ref{eq:rhothree}) suitably shifted to
vanish at the real critical point, describes the simulations for RR
networks much better than for WS networks. If, nevertheless, the
numerically calculated factor $g=1.0832$ is explicitly included in the
heterogeneous mean-field theory, Eq.~(\ref{eq:rhosup}), a very good
agreement is also found for WS networks.

\section{Conclusions}
\label{sec:conclusions}
The Watts-Strogatz model~\cite{watts1998collective} generates networks
that interpolate between regular lattices and random 
graphs allowing to adjust the clustering of the network with a rewiring 
probability $p$. The small-world property is always present on large 
networks for any $p>0$ while the clustering vanishes as $p\rightarrow 1$. 
We investigated the role of these structural properties on the dynamics of 
the contact processes, focusing on the critical properties of the ensuing 
absorbing state phase transition. We show that the critical 
exponents do not change when the rewiring probability and, consequently, 
the clustering coefficient are varied in a wide range. Thus, our results 
corroborate previous evidences that mean-field approximations
yield the correct critical exponents of statistical physics models running
on the top of WS networks for any finite rewiring
probability~\cite{barrat00,PhysRevE.65.066110}.

We also analyzed dimensionless moment ratios and show that universal
values are found independently of the rewiring probability $p>0$. The ratios are
equal to those obtained for annealed networks~\cite{cp_quenched}, for which they
are independent of the degree distribution $P(k)$, {differently
from the} distribution dependence observed in scale-free quenched
networks~\cite{cp_quenched}. Moreover, our results support the conjecture that
the moment ratios converge to the annealed values when
$\gamma\rightarrow\infty$~\cite{cp_quenched}.

Differently from the exponents and moment ratios, the critical points
strongly depend on the rewiring probability. Mean-field cluster
approximations (see appendixes~\ref{app:pair} and \ref{app:three}) yield a
very good prediction for critical point in the limit $p\rightarrow1$, but
fails for highly clustered networks. In fact, the existence of loops is
essential to the critical point determination as qualitatively clarified
in a modified pair mean-field theory (Appendix~\ref{app:pair}).

\section*{Acknowledgment}
Authors thank the financial support of the Brazilian agencies CNPq and
FAPEMIG. We are in debt with A. S. Mata by valuables 
comments on the manuscript.
\appendix
\section{Homogeneous pair-approximation}
\label{app:pair}
In a homogeneous pair-approximation, the states of the vertices do not
depend on their degrees. So, let us introduce the notation $[A]$ to
represent the probability that a vertex is in the state $A$, $[AB]$ that a
pair of nearest neighbors are in states $A$ and $B$, and $[ABC]$ the
generalization for a cluster of three vertices. Representing an occupied
vertex by $1$ and an empty one by $0$, we introduce the following
notation: $[1]=\rho$, $[0]=1-\rho$, $[10]=[01]=\phi$, $[11]=\psi$,
$[00]=\omega$. The normalization conditions $\psi+\phi=\rho$ and
$\psi+2\phi+\omega=1$ are derived straightforwardly. In addition, we
assume that the chance of two vertices to have one or more common
neighbors is negligible, which is an incorrect premise for highly
clustered networks. This approach is equivalent to the one used in
Ref.~\cite{Joo} for the susceptible-infected-recovered-susceptible epidemic model.
The following dynamical equations are obtained for CP dynamics in a
homogeneous network with a fixed vertex degree $k$
\begin{equation}
 \frac{d \rho}{d t} = -\rho+\lambda \phi
\label{eq:rho_2}
\end{equation}
and
\begin{equation}
 \frac{d \phi}{dt} = -\phi-\lambda\frac{\phi}{k}+\psi+
\lambda\frac{k-1}{k}[001]-\lambda\frac{k-1}{k}[101].
\label{eq:phi}
\end{equation}
The first three terms of Eq.~(\ref{eq:phi}) reckons the reactions
inside the pair that contribute to change the density $[01]$. The fourth
term increases the density $[01]$ due to the creation in the right vertex 
of an empty pair $00$ due to its other $k-1$ neighbors. Finally, 
the last one represents the creation in the empty vertex of a pair $01$.

We now truncate triplets in Eq.~(\ref{eq:phi}) using the
pair-approximation~\cite{Avraham92}
\[ [ABC]\approx\frac{[AB][BC]}{[B]},\]
resulting
\begin{equation}
 \frac{d \phi}{dt} = -\phi-\lambda\frac{\phi}{k}+\psi+
\lambda\frac{k-1}{k}\frac{\omega\phi-\phi^2}{1-\rho}.
\label{eq:phi_pair}
\end{equation}
Solving equations~(\ref{eq:rho_2}) and (\ref{eq:phi_pair}) 
in the stationary state ($d\rho/dt=d\phi/dt=0$) and using normalization
conditions, we obtain
\begin{equation}
\bar{\rho} = \frac{\lambda-\lambda_c}{1+\lambda-\lambda_c}
\label{eq:rhopair}
\end{equation}
with the critical point
\begin{equation}
\lambda_c = \frac{k}{k-1}.
 \label{eq:lbcpair}
\end{equation}

A simple approximation to clustered networks is derived assuming that 
the triplets form a loop with probability $c$, the clustering coefficient.
In the case of loops, only $k-2$ neighbors contribute to the
creation/annihilation of 01's in the last terms
of Eq.~(\ref{eq:phi}). We then rewrite Eq.~(\ref{eq:phi}) as
\begin{eqnarray}
 \frac{d \phi}{dt}&=&-\phi-\lambda\frac{\phi}{k}+\psi+
\lambda\left[\frac{k-1}{k}(1-c)+\frac{k-2}{k}c\right]\nonumber\\
&\times&([001]-[101]).
\label{eq:phi-cl}
\end{eqnarray}
Performing the pair-approximation we have
\begin{equation}
\lambda_c=\frac{k}{k-1-c/2},
 \label{eq:lbcpair-cl}
\end{equation}
which qualitatively explains the critical points increasing for larger
clustering. 

The clustering coefficient of WS model as a function of $p$ is well 
described by~\cite{barrat00}
\begin{equation}
c(p)=\frac{3}{4}\frac{(K-2)}{(K-1)}(1-p)^3.
\label{eq:c_p}
\end{equation}
Using the clustering coefficient given by Eq.~(\ref{eq:c_p}), one obtains
from the Eq.~(\ref{eq:lbcpair-cl}) $\lambda_c=1.2742$ and  $1.2548$ for  
$p=0.01$ and $p=0.10$, respectively, which are quite far from the 
transition points observed in simulations of highly clustered networks 
given in Table~\ref{tab:mom}.

\section{Homogeneous three-site approximation}
\label{app:three}
Lets us introduce the notation for triplets: $a=[001]=[100]$,
$b=[011]=[110]$, $c=[111]$, $d=[000]$, $e=[010]$, and $f=[101]$. We have 
that the following independent relations hold: $a+f=\phi$, $b+c=\psi$,
$e+b=\phi$, $d+a=\omega$. We need two more independent equations. 
We have chosen to write them for $a$ and $e$. The resulting
dynamical equations are
\begin{eqnarray}
\frac{da}{dt}&=&-\left(1+\frac{\lambda}{k}\right)a+b+f
+\lambda\frac{k-1}{k}([0001]-[1001])\nonumber\\
&-&\lambda\frac{k-2}{k}[1001]
 \label{eq:a}
\end{eqnarray}
and
\begin{equation}
\frac{d e}{dt}=-\left(1+\frac{2\lambda}{k}\right)e+2b
+\lambda\frac{k-2}{k}[1000]-2\lambda\frac{k-1}{k}[1010].
 \label{eq:e}
\end{equation}
The first three terms of Eq.~(\ref{eq:a}) represent the reactions inside
the triplet $001$, the forth term represents the events caused by
the creation due to the $k-1$ nearest neighbors of the rightmost and
leftmost vertices of the triplets $000$ and $001$, respectively, and the
last one reckons the creation due to the $k-2$ nearest neighbors of the
middle vertex of the triplet $001$. Similar interpretations hold for
Eq.~(\ref{eq:e}).

We now replace quadruplets using the cluster
approximation~\cite{Avraham92}
\begin{equation}
[ABCD] \approx \frac{[ABC][BCD]}{[BC]}
 \label{eq:quadruplets}
\end{equation}
to obtain
\begin{equation}
\frac{da}{dt}=-\left(1+\frac{\lambda}{k}\right)a+b+f
+\lambda\frac{k-1}{k}\frac{da}{\omega}-2\lambda\frac{2k-3}{k}
\frac{a^2}{\omega}
\label{eq:a2}
\end{equation}
and
\begin{equation}
\frac{d e}{dt}=-\left(1+\frac{2\lambda}{k}\right)e+2b
+\lambda\frac{k-2}{k}\frac{ad}{\omega}-2\lambda\frac{k-1}{k}\frac{fe}{\phi}.
 \label{eq:e2}
\end{equation}

Solving Eqs.~(\ref{eq:rho_2}), (\ref{eq:phi}),
(\ref{eq:a2}) and (\ref{eq:e2}) with the normalization conditions, the 
stationary density is
\begin{equation}
\bar{\rho} = \frac{(k-1)(3k-4)(\lambda-\lambda_c)(\lambda+\lambda_c
-\frac{2k}{3k-4})}
{\lambda(k-1)[(3k-4)\lambda+(k-4)]-k},
 \label{eq:rhothree}
\end{equation}
where the critical point is given by
\begin{equation}
\lambda_c = \frac{k+2\sqrt{k^2-k}}{3k-4}.
 \label{eq:lbcthree}
\end{equation}
It is worth mentioning that independently of the apparently very
different forms of Eqs.~(\ref{eq:rhopair}) and (\ref{eq:rhothree})
they are almost indistinguishable in plots $\bar{\rho}$ versus
$\lambda-\lambda_c$.
%
\bibliographystyle{epj}

%
\end{document}